# On Von Schelling Formula for the Generalized Coupon Collector Problem

**Christian BERTHET**
STMicroelectronics, Grenoble, France,

**Abstract**.  This paper gives an algebraic proof of the correctness of Von Schelling formula for the probability of the coupon collector problem waiting time for non-uniform distributions and partial collections. It introduces a theorem on sums of powers of subset probabilities which to our knowledge is new. A set of binomial coefficients is used as a basis for decomposition of these sums of powers.

**Address all correspondence to:** BERTHET Christian; E-mail: Christian.berthet@st.com

## Introduction

The Coupon Collector Problem (CCP) is a totem pole in the field of combinatorial problems and has many implications in different domains, e.g., the exact Miss rate of LRU caches.
In CCP, a set {1,..n}, n>1, of distinct objects (coupons, items...) is sampled with replacement by a collector. This random process, frequently labelled Independent Reference Model (IRM), is independent of all past events. Each drawing produces item 'i' from the reference set with a given probability, which is unequal in the generalized CCP. The problem is to know how many trials ('expected waiting time') are needed before one has collected n items for the complete collection, and a number c, $1 \leq c < n$, of items for a partial collection.
A somewhat obscure formula exists since Von Schelling seminal paper in 1954 [VonSchelling54] that gives the solution to this problem. However, to our knowledge it has never been proved that this mathematical expression is correct.
In this paper we focus on showing that Von Schelling formula is coherent with the obvious statement that no one can collect c coupons in less than c trials. It appears that this proof is not trivial and exhibit some combinatorial difficulties.
This paper proposes a purely algebraic approach which lead us to additional material in the form of a theorem on a decomposition property applied to sums of powers of subset probabilities.

## Von Schelling Formula

*Notation*

We assume a probability law with general distribution $\{p_i\}$, $i \in \{1..n\}$, $n > 1$, $1 > p_i > 0$, $\sum_{i=1}^{n} p_i = 1$.

This probability law is often called 'popularity'. For a subset J of the reference set {1,..,n}, we note $P_J$ the subset probability, i.e. the sum of probabilities of the elements of J: $P_J = \sum_{i \in J} p_i$. We note $\sum_{|J|=j}$ the summation operator over all subsets J of size j where j is: 0≤j≤n. In particular, it stands that $\sum_{|J|=0} P_J = 0$, $\sum_{|J|=n} P_J = 1$, and $\sum_{|J|=j} 1 = \binom{n}{j}$. By convention: $\sum_{|J|=0} 1 = 1$. Implicitly and unless otherwise specified, the summation operator $\sum_J$ applies to all subsets of the reference set. Also, in the sequel, it goes without saying that, for a and b integers, $\binom{a}{b}$ is equal to 1 when a=b, else it is null if b>a or if any of them is negative.

*Some History*

As far back as 1954 (and even mentioning a previous work done in 1934) Von Schelling published formulas for probability, expectation and variance of the CCP waiting time for a general popularity. His work analyzes the probability that the $m^{th}$-last event (missing item) is observed at a given time. In this definition, the probability for the last event (i.e., m=1) is that of the full collection.
Although the probability density function (pdf) formula (2) page 307 lacks a binomial coefficient in the last term of the summation (it can be easily devised from formula (4) in the subsequent page 308), Von Schelling formula essentially gave the core of the solution of CCP for a general popularity.
In the sequel, we use Von Schelling formula transformed by noticing that rank m of the $m^{th}$-last event is such that: m=n-c+1 (where n is the reference size and c the sub-collection size), and using a simpler and more modern notation of summation over subsets.
Then, denoting by $T_c$ the random variable ("waiting Time") indicating that a sub-collection of size c from the set of n items is completed in k trials, Von Schelling pdf formula is:

$$\Pr[T_c = k] = \sum_{j=0}^{c-1} (-1)^{c-1-j} \binom{n-j-1}{n-c} \sum_{|J|=j} P_J^{k-1}(1 - P_J)$$

It is worth noting that the number of trials k, is assumed to be greater or equal to n-m+1=c at the end of formula (2) in Von Schelling paper.
The Cumulative probability (CDF) can be directly derived from pdf formula and is:

$\Pr[T_c \leq k] = \sum_{j=0}^{c-1} (-1)^{c-1-j} \binom{n-j-1}{n-c} \sum_{|J|=j} (1 - P_J^k)$, and bearing in mind the combinatorial identity (see Appendix 1): $\sum_{j=0}^{c-1} (-1)^{c-1-j} \binom{n-j-1}{n-c} \binom{n}{j} = 1$, this is: $\Pr[T_c \leq k] = 1 - \sum_{j=0}^{c-1} (-1)^{c-1-j} \binom{n-j-1}{n-c} \sum_{|J|=j} P_J^k$.

Regarding this CDF expression, Von Schelling indicates that, when the number of trials goes to infinity, it equals to 1, which stems from the fact that the size of any subset J is always less than c (which is less or equal to n) and thus $P_J^k$ tends to 0 when exponent k increases.

Equivalently, using the Complementary CDF (CCDF): $\Pr[T_c > k] = \sum_{j=0}^{c-1}(-1)^{c-1-j}\binom{n-j-1}{n-c}\sum_{|J|=j}P_J^k$,

relation $\Pr[T_c > 0] = \sum_{j=0}^{c-1}(-1)^{c-1-j}\binom{n-j-1}{n-c}\sum_{|J|=j}P_J^0 = \sum_{j=0}^{c-1}(-1)^{c-1-j}\binom{n-j-1}{n-c}\binom{n}{j} = 1$ is easily shown by setting k=0 in the relation in Appendix 1.

However, the fact that the mathematical expression of $\Pr[T_c = k]$ has to be 0 when k<c is not explicitly stated in Von Schelling paper. More precisely, speaking of a specific sum, he mentioned that "Indeed it can be proved that this sum equals 1". This sum is, using our notation: $\sum_{j=0}^{c-1}(-1)^{c-1-j}\binom{n-j-1}{n-c}\sum_{|J|=j}P_J^{c-1}$, i.e $\Pr[T_c > c-1]$. Unfortunately, he did not give a hint of his proof.

In our mind, even this proof is not sufficient to ensure the correctness of the probability formula. In this paper, we focus on proving that $\Pr[T_c > k] = 1$, for any k: 0≤k<c≤n. To be complete, it should be needed to show also that $\Pr[T_c = k] > 0$, $\forall k \geq c$.

*Particular case of a uniform popularity*

The probability for a uniform popularity has a simple expression with Stirling numbers of the second order: $\Pr[T_c = k] = \frac{n!S(k-1,c-1)}{(n-c)!n^k}$ (for a complete collection this even reduces to $\Pr[T_n = k] = \frac{n!S(k-1,n-1)}{n^k}$, formula (3) of [Boneh97]). Thanks to the properties of Stirling numbers of 2$^{nd}$ order, this expression verifies: $\Pr[T_c = k] = 0$, $\forall k < c$ and $\Pr[T_c = k] > 0$, $\forall k \geq c$. It is a unimodal distribution with a maximum, and $\lim_{k \to \infty} \Pr[T_c = k] = 0^+$, which stems from the asymptotic approximation $S(n,d) \approx \frac{d^n}{d!}$ (http://dlmf.nist.gov/26.8#vii), when n→∞, hence

$\Pr[T_c = k] \approx \binom{n-1}{c-1}\left(\frac{c-1}{n}\right)^{k-1} \to 0^+$ when k→∞.

Thus, a legitimate question is: does the same hold for any general popularity?

*Flajolet & al. approach*

In 1992, Flajolet, Gardy and Thimonier proposed a completely new point of view, introducing generating functions [Flajolet92]. See also [Boneh97]. Using this mathematical apparatus, CCP probability is: $\Pr[T_c \leq k] = \sum_{c \leq |J| \leq n} k! [z^k] \prod_{i \in J}(e^{p_i z} - 1)$, where [$z^k$] denotes the coefficient of $z^k$ in the polynomial. Since $e^{p_i z} - 1 = \sum_{n>0} \frac{(p_i z)^n}{n!}$ is a polynomial of degree at least 1, this probability is null for k<c, which immediately proves Von Schelling claim. Other direct consequences of this formulation are: $\Pr[T_c = c] = c! \sum_{|J|=c} \prod_{i \in J} p_i$ and $\Pr[T_n = n] = n! \prod_{i=1}^{n} p_i$.

However this powerful approach was not available to Von Schelling at the time of his paper. So we can expect his 1954 proof was using only pure algebraic techniques. In the sequel, we prove using solely algebraic means that $\Pr[T_c > k] = 1$, for any k: 0≤k<c≤n, which implies for any k: 0<k<c≤n: $\Pr[T_c = k] = 0$.

*Other Works*

It is sometimes claimed that Rubin was the first to give a formula [Rubin65] of the pdf probability for an incomplete collection with a uniform popularity. In fact, his formula obviously can be derived from Von Schelling one.

More recently, Brown, Peköz and Ross gave the following pdf formula (proposition 1b/ p. 222 [Brown08]) for the complete collection where n is the number of coupons, J is a subset of size j of {1,..,n} and k the number of trials: $\sum_{j=1}^{n} (-1)^{j-1} \sum_{|J|=j} q_J^{k-1} p_J$ with $p_J = \sum_{j \in J} p_j$ and $q_J = 1 - p_J$.

This expression is assumed to be null for k<n. There again, this can be derived with a suitable index change from Von Schelling formula for a complete collection. In this work, there is no formulation in the case of an incomplete collection.

Using a model based on Markov chains [Anceaume14], Anceaume, Busnel and Sericola give a general formula for the probability in case of an incomplete set. They also extend the CCP with a so-called null coupon but we ignore it here. The CCDF probability of the waiting time needed for an incomplete collection of size c among n possible in k trials is (Theorem 2 p.4) for

1<=c<=n: $P(T_c > k) = \sum_{j=0}^{c-1} (-1)^{c-1-j} \binom{n-j-1}{n-c} \sum_{|J|=j} P_J^k$. As previously mentioned, this relation is a re-writing of Von Schelling formula. The correctness of this formula is claimed by construction of the Markov chain [Anceaume14, Corollary 3 pg.5].

**Correctness of the probability formula**

We have seen previously that CCDF verifies $\Pr[T_c > 0] = 1$. This follows from the relation proven in Appendix 1 Preliminary Combinatorial Identities. Similarly, using the same identity and anticipating relation $\sum_{|J|=j} P_J = \binom{n-1}{j-1}$ proven in Appendix 2 relation 1, we have:

$\Pr[T_c > 1] = \sum_{j=0}^{c-1} (-1)^{c-1-j} \binom{n-j-1}{n-c} \binom{n-1}{j-1} = 1$.

However, proving that the same result holds for any k, k<c≤n, reveals much more complex.

We first give the following Theorem.

*Theorem 1*

*For all k, 0<k<n, and j, 0≤j≤n, there exists a set of coefficients ('weights') $\alpha_{k,u}$, 1≤u≤k, such that:*

$$\boxed{\sum_{|J|=j} P_J^{\ k} = \sum_{u=1}^{k} \binom{n-k}{j-u} \alpha_{k,u}}$$

*where $\alpha_{k,u}$ are independent of j (i.e. depend only on n, k and the popularity) and $\alpha_{k,k} = 1$.*

Note that Theorem 1 could be trivially extended to k=0 (with index u starting at 0), since $\sum_{|J|=j} P_J^{\ 0} = \sum_{u=0}^{0} \binom{n}{j-u} \cdot \alpha_{0,u} = \binom{n}{j} \cdot \alpha_{0,0}$ with $\alpha_{0,0} = 1$, however this does not bring much to the computation and it implies handling, for any k>0, $\alpha_{k,0} = 0$.

The proof of Theorem 1 is given in Appendix 2. Prior to this proof, in Appendix 2 as well, the three following relations are proven. Respectively, Relation 1, Relation 4 and Relation 7:

$$\sum_{|J|=j} P_J = \binom{n-1}{j-1}; \sum_{|J|=j} P_J^{\ 2} = \binom{n-2}{j-2} + \binom{n-2}{j-1} \sum_{l=1}^{n} p_l^{\ 2}; \sum_{|J|=j} P_J^{\ 3} = \binom{n-3}{j-3} + \binom{n-3}{j-2}\left(3\sum_{l=1}^{n} p_l^{\ 2} - \sum_{l=1}^{n} p_l^{\ 3}\right) + \binom{n-3}{j-1} \sum_{l=1}^{n} p_l^{\ 3}.$$

For the general case of exponent k, proof is given in Relation 8 and 9.

Let us note that theorem is obvious for j=1 since $\sum_{|J|=1} P_J^{\ k} = \sum_{l=1}^{n} p_l^{\ k}$ which can be written $\sum_{u=1}^{k} \binom{n-k}{1-u} \cdot \alpha_{k,u}$ with $\alpha_{k,1} = \sum_{l=1}^{n} p_l^{\ k}$, and implicitly $\alpha_{k,k} = 1$, k>1. More generally, when j<k, binomial coefficients for u above j are null, and $\alpha_{k,k}$ is implicitly equal to 1.

Also, since $\sum_{|J|=n} P_J^{\ k} = 1$, expression for j=n is: $\sum_{u=1}^{k} \binom{n-k}{n-u} \cdot \alpha_{k,u} = \alpha_{k,k} = 1$.

We can note the analogy to the use of binomial coefficients as a basis for the space of polynomials. Indeed $\sum_{|J|=j} P_J^{\ k}$ is a polynomial of multiple variables and whose sum of the degrees of the variables is equal to k.

We can now proceed with the proof of the general CCDF formula. Prior to that, we prove a more restricted but interesting result.

*Corollary 1*
$$\sum_{0\leq |J|\leq n}(-1)^{|J|}P_J^k = 0, \quad 0\leq k < n$$

Proof is direct using decomposition: $\sum_{j=0}^{n}(-1)^j\sum_{|J|=j}P_J^k = \sum_{j=0}^{n}(-1)^j\sum_{u=1}^{k}\binom{n-k}{j-u}\cdot \alpha_{k,u}$, and noting that the inner summation of $\sum_{u=1}^{k}\alpha_{k,u}\sum_{j=0}^{n}(-1)^j\binom{n-k}{j-u}$ is null for u≤k.

Note that for a uniform distribution, corollary 1 lends to the well-known identity: $\sum_{j=0}^{n}(-1)^j\binom{n}{j}j^k = 0$, 0≤k<n, [Boyadzhiev12]. Hence, it can be seen as an extension to non-uniform distributions of this identity.

The general relation for the probability of incomplete collections follows.

*Corollary 2*
$$\sum_{0\leq |J|\leq c-1}(-1)^{c-1-|J|}\binom{n-|J|-1}{n-c}P_J^k = 1, \quad 0\leq k < c \leq n$$

When c=n, this leads to: $\sum_{j=0}^{n-1}(-1)^{n-1-j}\sum_{|J|=j}P_J^k = 1$, 0≤k<n, equivalent to corollary 1.

Again, proof is direct using decomposition: $\sum_{|J|=j}P_J^k = \sum_{u=1}^{k}\binom{n-k}{j-u}\cdot \alpha_{k,u}$, with $\alpha_{k,u}$ coefficients independent of j, and $\alpha_{k,k}=1$, it suffices to show that: $\sum_{j=0}^{c-1}(-1)^{c-1-j}\binom{n-j-1}{n-c}\binom{n-k}{j-k} = 1$ and $\sum_{j=0}^{c-1}(-1)^{c-1-j}\binom{n-j-1}{n-c}\binom{n-k}{j-u} = 0$ for 1≤u<k. These two combinatorial identities are proven in Appendix 1.

This concludes the proof that Von Schelling probability verifies: $\Pr[T_c > k] = 1$, for k: 0≤k<c≤n.

There is an analytic expression of $\alpha_{k,u}$ coefficients when the popularity is uniform. This expression is given in Appendix 3: $\alpha_{k,u} = \frac{1}{n^k}\sum_{i=1}^{n}S(k,i)i!\binom{k-i}{u-i}\binom{n}{i}$.

**Conclusion**

Using algebraic means we have shown the correctness of the CCP probability definition for a partial collection assuming a general popularity, initially stated by Von Schelling seminal paper. It can be argued that the exercise is a waste of time, stating the obvious that one cannot collect a given number of coupons in a smaller number of trials.

However the outcome is satisfactory for the mind and the analysis leads to a new decomposition property of powers of subset probabilities that can be profitable elsewhere.
The author can only hope a faster and less tedious proof of Theorem 1 will be devised.

**Acknowledgments**

The author thanks Batominovski for his helpful proof for a variant of the decomposition for uniform distributions (http://math.stackexchange.com/questions/1841501/decomposition-of-binom-n-j-1jk/1841530#1841530).

**Post-scriptum**

During the finalization phase of this document, a major contribution was made by Marko Riedel (http://math.stackexchange.com/questions/2132016/generalization-of-the-well-known-combinatorial-identity-sum-j-0n-1j-n) who gave an analytic form of decomposition of $\sum_{|J|=j} P_J^k$ when j≥k (in fact j>k, because the analytic form lends to a tautology for j=k).
Although this form is not practical for proving Corollary 2 (for which all of 0≤j<c is considered while 0<k<c), we show in Appendix 4 that his decomposition can be ported to ours. Still, this decomposition, in addition to j>k, requires that Corollary 2 must hold for k=c-1 (|Hence Corollary 2 then must be proven by other means…)
However, a potentially non-negligible application (derived in Appendix 5) is that, when j is large and k is small (j>>k), this decomposition leads to a much more efficient way to compute a sum of powers of j-size subset probabilities as a sum of powers of q-size subset probabilities, for q from 1 to k.

**Appendix 1: Preliminary Combinatorial identity**
$$\sum_{i=0}^{c-1}(-1)^{c-1-i}\binom{n-i-1}{n-c}\binom{n-k}{i-u} = \begin{cases} 1 & \text{for } 0\leq u=k<c\leq n \\ 0 & \text{for } 0\leq u<k<c\leq n \end{cases}$$

The following relation, obtained by iterated integration by parts, holds for a≥0, b≥0:
$$\int_0^1 x^a(1-x)^b\,dx = \frac{a!b!}{(a+b+1)!} = \frac{1}{b+1}\binom{a+b+1}{b+1}^{-1}.$$ By integrating on x the binomial identity:

$$x^a(1-x)^b = \sum_{i=0}^b \binom{b}{i}(-1)^i x^{a+i},$$ one has: $\sum_{i=0}^b \binom{b}{i}(-1)^i \frac{b+1}{a+i+1}\binom{a+b+1}{b+1} = 1$, or

$$\sum_{i=0}^b (-1)^i \binom{b+1}{i+1}\binom{a+b+1}{b+1}\frac{i+1}{a+i+1} = 1.$$ Then, with index change (b-i):

$$\sum_{i=0}^b (-1)^{b-i}\binom{b+1}{i}\binom{a+b+1}{b+1}\frac{b-i+1}{a+b-i+1} = 1,$$ subset-of-a-subset identity:

$$\sum_{i=0}^b (-1)^{b-i}\binom{a+b+1}{i}\binom{a+b+1-i}{b+1-i}\frac{b-i+1}{a+b-i+1} = 1$$ and simplification: $\sum_{i=0}^b (-1)^{b-i}\binom{a+b+1}{i}\binom{a+b-i}{a} = 1$.

Case k=u=0 is trivial by setting a=n-c and b=c-1 (both ≥0): $\sum_{i=0}^{c-1}(-1)^{c-1-i}\binom{n}{i}\binom{n-1-i}{n-c} = 1$

With a=n-c and b=c-1-k (both ≥0 since 0≤k<c≤n): $\sum_{i=0}^{c-1-k}(-1)^{c-1-k-i}\binom{n-k}{i}\binom{n-1-k-i}{n-c} = 1$

With index change (i+k): $\sum_{i=k}^{c-1}(-1)^{c-1-i}\binom{n-k}{i-k}\binom{n-1-i}{n-c} = 1$, and noting that the first binomial is null for i<k, yields the desired result for u=k: $\sum_{i=0}^{c-1}(-1)^{c-1-i}\binom{n-k}{i-k}\binom{n-1-i}{n-c} = 1$.

Note the particular case for 0≤u=k<c=n: $\sum_{i=0}^{n-1}(-1)^{n-1-i}\binom{n-k}{i-k} = 1$ or: $\sum_{i=0}^n (-1)^i \binom{n-k}{i-k} = 0$.

We can now prove that $\sum_{i=0}^{c-1}(-1)^{c-1-i}\binom{n-i-1}{n-c}\binom{n-k}{i-u} = 0$, for 0≤u<k<c≤n, by noting it is direct for k-u=1

since $\binom{n-k}{i-(k-1)} = \binom{n-(k-1)}{i-(k-1)} - \binom{n-k}{i-k}$. And, if this is true for a value v=k-u less than k, it is true for

v+1=k-(u-1), using $\binom{n-k}{i-(u-1)} = \binom{n-(k-1)}{i-(u-1)} - \binom{n-k}{i-u}$.

A direct consequence, for c=n, is $\sum_{i=0}^{n-1}(-1)^i \binom{n-k}{i-u} = 0$, for 0≤u<k<n. Since n-u>n-k, ranges of u and index summation can be extended: $\sum_{i=0}^n (-1)^i \binom{n-k}{i-u} = 0$, 0≤u≤k<n.

**Appendix 2: Relations on Sums of Subset Probabilities**

*Relation 1*
$$\boxed{\sum_{|J|=j} P_J = \binom{n-1}{j-1}, \quad 0 \leq j \leq n}$$

Since $\binom{n}{j}\frac{j}{n} = \binom{n-1}{j-1}$, $0 \leq j \leq n$, relation means that the sum of probabilities of all subsets of size j is the same as for a uniform distribution: $\sum_{|J|=j} P_J = \sum_{|J|=j} \frac{j}{n}$.

In other words, whatever the considered distribution, the average value of $P_J$ for each of the subsets is j/n, i.e., the subset probability of the uniform distribution.

Proof is direct for j=0 since $\sum_{|J|=0} P_J = 0$. Also for j=1: $\sum_{|J|=1} P_J = \sum_{1 \leq i \leq n} p_i = 1$; For j=2, we have:

$$\sum_{|J|=2} P_J = \sum_{1 \leq i < j \leq n}(p_i + p_j) = \sum_{1 \leq i < j \leq n} p_i + \sum_{1 \leq i < j \leq n} p_j = \sum_{1 \leq i \leq n} p_i(n-i) + \sum_{1 \leq j \leq n} p_j(j-1) = \sum_{1 \leq i \leq n} p_i(n-1) = n-1$$

with a suitable variable renaming.

In the general case, let J={$i_1 < i_2 < .. < i_j$} a subset of {1,..,n} of size j.

$$\sum_{|J|=j} P_J = \sum_{1 \leq i_1 < i_2 < .. < i_j \leq n}(p_{i_1} + p_{i_2} + .. + p_{i_j}) = \sum_{1 \leq i_1 < i_2 < .. < i_j \leq n} p_{i_1} + .. + \sum_{1 \leq i_1 < i_2 < .. < i_j \leq n} p_{i_j} = \sum_{k=1}^{j} \sum_{1 \leq i_k \leq n} p_{i_k} \binom{i_k - 1}{k-1}\binom{n - i_k}{j-k}$$

This quantity represents for each possible ranking k, 1<=k<=j, and each possible position $i_k$ of the ranking k in the set {1,..,n} according to the ordering ($1 \leq i_1 < i_2 < .. i_j \leq n$), the number of ways to choose a subset of size (k-1) among ($i_k$-1) possible candidates below that element, times the number of ways to choose a subset of size (j-k) among (n- $i_k$) candidates above that element.

After renaming each variable (since they belong to independent parts of the second sum) and commuting the summation order:

$$\sum_{|J|=j} P_J = \sum_{1 \leq i \leq n} p_i \sum_{k=1}^{j} \binom{i-1}{k-1}\binom{n-i}{j-k} = \sum_{1 \leq i \leq n} p_i \sum_{k=1}^{j} \binom{i-1}{k-1}\binom{n-1-(i-1)}{j-1-(k-1)} = \sum_{1 \leq i \leq n} p_i \sum_{k=0}^{j-1} \binom{i-1}{k}\binom{n-1-(i-1)}{j-1-k}.$$

At this point, Chu-Vandermonde identity steps in: $\sum_{j=0}^{k} \binom{m}{j}\binom{n-m}{k-j} = \binom{n}{k}$. It comes from identity

$[x^k] ((1 + x)^m (1 + x)^{n-m}) = [x^k] (1 + x)^n$, $\forall$ k, m integers $0 \leq k \leq n$, and $0 \leq m \leq n$.

This yields: $\sum_{|J|=j} P_J = \sum_{1 \leq i \leq n} p_i \binom{n-1}{j-1} = \binom{n-1}{j-1}$. QED.

Consequences of this relation are: $\sum_{|J|=j}(1-P_J) = \binom{n}{j} - \binom{n-1}{j-1} = \binom{n-1}{j} = \sum_{|J|=j+1} P_J$,

hence: $\sum_{|J|=j+1} P_J + \sum_{|J|=j} P_J = \binom{n}{j}$. Also $\sum_{|J|=n-j} P_J = \sum_{|J|=j+1} P_J$.

We introduce an additional notation in the subset summation, where a specific element of {1,..,n}, does not belong to J.

*Relation 2*
$$\sum_{\substack{|J|=j \\ l \notin J}} P_J = (1-p_l)\binom{n-2}{j-1}, \quad 1 \le j \le n-1$$

It is direct that $\sum_{\substack{|J|=1 \\ l \notin J}} P_J = (1-p_l)$. We use induction on j:

$$\sum_{\substack{|J|=j \\ l \in J}} P_J = \sum_{\substack{|J|=j-1 \\ l \notin J}} p_l + \sum_{\substack{|J|=j-1 \\ l \notin J}} P_J = p_l \sum_{\substack{|J|=j-1 \\ l \notin J}} 1 + \sum_{\substack{|J|=j-1 \\ l \notin J}} P_J = p_l \binom{n-1}{j-1} + (1-p_l)\binom{n-2}{j-2}$$

Also, $\sum_{|J|=j} P_J = \sum_{\substack{|J|=j \\ l \in J}} P_J + \sum_{\substack{|J|=j \\ l \notin J}} P_J$ hence: $\sum_{\substack{|J|=j \\ l \notin J}} P_J = \binom{n-1}{j-1} - p_l\binom{n-1}{j-1} - (1-p_l)\binom{n-2}{j-2} = (1-p_l)\binom{n-2}{j-1}$. QED.

Consequently: $\sum_{\substack{|J|=j \\ l \in J}} P_J = \binom{n-2}{j-2} + p_l\binom{n-2}{j-1}$. In particular: $\sum_{\substack{|J|=1 \\ l \in J}} P_J = p_l$.

We introduce higher powers of the subset probabilities:

*Relation 3*
$$\boxed{\sum_{|J|=j} P_J^k(1-P_J) = \sum_{l=1}^n \left( p_l \sum_{|J|=j, l \notin J} P_J^k \right), \forall k \ge 0}$$

This stems from: $\sum_{|J|=j} P_J^k(1-P_J) = \sum_{|J|=j} P_J^k (\sum_{l=1, l\notin J}^n p_l) = \sum_{l=1}^n \sum_{|J|=j, l\notin J} P_J^k p_l = \sum_{l=1}^n \left( p_l \sum_{|J|=j, l\notin J} P_J^k \right)$. QED.

*Relation 4 (ABS Lemma)*
$$\sum_{|J|=j} P_J^{k+1} = \sum_{l=1}^{n}\left(p_l \sum_{\substack{|J|=j \\ l \in J}} P_J^{k}\right), \quad 1 \leq j \leq n, \; 0 \leq k$$

This lemma is proved in [Anceaume 14] (pg. 2, with a=0 and rewritten with our notation) by induction on the size n of the reference set. Another proof of this lemma using Relation 3 is:

$$\sum_{|J|=j} P_J^{k+1} = \sum_{|J|=j} P_J^{k} - \sum_{|J|=j} P_J^{k}(1-P_J) = \left(\sum_{l=1}^{n} p_l\right)\sum_{|J|=j} P_J^{k} - \sum_{l=1}^{n}\left(p_l \sum_{\substack{|J|=j \\ l \notin J}} P_J^{k}\right) = \sum_{l=1}^{n}\left(p_l \sum_{\substack{|J|=j \\ l \in J}} P_J^{k}\right). \quad \text{QED.}$$

This lemma allows for the computation of sums of subset probabilities at the power of two:

$$\sum_{|J|=j} P_J^{2} = \sum_{l=1}^{n}\left(p_l \sum_{\substack{|J|=j \\ l \in J}} P_J\right) = \sum_{l=1}^{n}\left(p_l\left(\binom{n-2}{j-2} + p_l\binom{n-2}{j-1}\right)\right) = \binom{n-2}{j-2} + \binom{n-2}{j-1}\sum_{l=1}^{n} p_l^{2}.$$

In particular $\sum_{|J|=1} P_J^{2} = \sum_{l=1}^{n} p_l^{2}$ and $\sum_{|J|=n} P_J^{2} = 1$. Result can be checked directly for j=2:

$$\sum_{|J|=2} P_J^{2} = \sum_{1 \leq i < j \leq n}(p_i + p_j)^{2} = \sum_{i=1}^{n} p_i^{2}(n-i) + \sum_{j=1}^{n} p_j^{2}(j-1) + \sum_{1 \leq i < j \leq n} 2p_i p_j = 1 + (n-2)\sum_{i=1}^{n} p_i^{2}$$

Applied to a uniform distribution, the following combinatorial identity results:

$$\binom{n}{j}\left(\frac{j}{n}\right)^{2} = \binom{n-2}{j-2} + \binom{n-2}{j-1}\left(\frac{1}{n}\right), \forall j : 0 \leq j \leq n.$$

*Relation 5*
$$\sum_{|J|=j} P_J(1-P_J) = \binom{n-2}{j-1}\left(1 - \sum_{l=1}^{n} p_l^{2}\right)$$

$$\sum_{|J|=j} P_J(1-P_J) = \sum_{|J|=j} P_J - \sum_{|J|=j} P_J^{2} = \binom{n-1}{j-1} - \binom{n-2}{j-2} + \binom{n-2}{j-1}\sum_{l=1}^{n} p_l^{2} = \binom{n-2}{j-1}\left(1 - \sum_{l=1}^{n} p_l^{2}\right). \quad \text{QED}$$

Interesting form of Relation 5 is: $\sum_{|J|=j} P_J(1-P_J) = \binom{n-2}{j-1} \cdot \sum_{|J|=1} P_J(1-P_J)$.

A consequence for j<n is also: $\sum_{|J|=j} P_J(1-P_J) = \sum_{|J|=j+1} P_J(P_J - \sum_{l=1}^{n} p_l^{2})$.

This comes from $\sum_{|J|=j} P_J^{2} + \sum_{|J|=j+1} P_J^{2} = \binom{n-1}{j-1} + \binom{n-1}{j}\sum_{l=1}^{n} p_l^{2} = \sum_{|J|=j} P_J + \sum_{|J|=j+1} P_J \cdot \sum_{l=1}^{n} p_l^{2}$.

*Relation 6*
$$\sum_{\substack{|J|=j \\ l \in J}} P_J^2 = p_l^2\left(\binom{n-3}{j-1}-\binom{n-3}{j-2}\right)+2p_l\binom{n-3}{j-2}+\binom{n-3}{j-3}+\binom{n-3}{j-2}\sum_{m=1}^n p_m^2$$

Proof is as follows: $\sum_{\substack{|J|=j \\ l \in J}} P_J^2 = \sum_{\substack{|J|=j-1 \\ l \notin J}}(p_l+P_J)^2 = p_l^2 \sum_{\substack{|J|=j-1 \\ l \notin J}} 1 + 2p_l \sum_{\substack{|J|=j-1 \\ l \notin J}} P_J + \sum_{\substack{|J|=j-1 \\ l \notin J}} P_J^2$

Then: $\sum_{\substack{|J|=j \\ l \in J}} P_J^2 + \sum_{\substack{|J|=j-1 \\ l \in J}} P_J^2 = p_l^2 \sum_{\substack{|J|=j-1 \\ l \notin J}} 1 + 2p_l \sum_{\substack{|J|=j-1 \\ l \notin J}} P_J + \sum_{|J|=j-1} P_J^2$.

And using previous results:

$$\sum_{\substack{|J|=j \\ l \in J}} P_J^2 + \sum_{\substack{|J|=j-1 \\ l \in J}} P_J^2 = p_l^2\binom{n-1}{j-1}+2p_l(1-p_l)\binom{n-2}{j-2}+\left[\binom{n-2}{j-3}+\binom{n-2}{j-2}\sum_{m=1}^n p_m^2\right]$$

$$= p_l^2\left(\binom{n-2}{j-1}-\binom{n-2}{j-2}\right)+2p_l\binom{n-2}{j-2}+\left[\binom{n-2}{j-3}+\binom{n-2}{j-2}\sum_{m=1}^n p_m^2\right]$$

Similarly down to j=1, multiplying each row by $(-1)^j$ and adding the rows lead to:

$(-1)^j \sum_{\substack{|J|=j \\ l \in J}} P_J^2 =$

$$= p_l^2\left(\sum_{k=1}^j(-1)^k\binom{n-2}{k-1}-\sum_{k=1}^j(-1)^k\binom{n-2}{k-2}\right)+2p_l\sum_{k=1}^j(-1)^k\binom{n-2}{k-2}+\sum_{k=1}^j(-1)^k\binom{n-2}{k-3}+\sum_{k=1}^j(-1)^k\binom{n-2}{k-2}\sum_{m=1}^n p_m^2$$

$$= p_l^2\left(\sum_{k=0}^{j-1}(-1)^{k+1}\binom{n-2}{k}-\sum_{k=-1}^{j-2}(-1)^{k+2}\binom{n-2}{k}\right)+2p_l\sum_{k=-1}^{j-2}(-1)^{k+2}\binom{n-2}{k}+\sum_{k=-2}^{j-3}(-1)^{k+3}\binom{n-2}{k}+\sum_{k=-1}^{j-2}(-1)^{k+2}\binom{n-2}{k}\sum_{m=1}^n p_m^2$$

Finally using the relation $\sum_{k=0}^j \binom{n}{k}(-1)^k = (-1)^j\binom{n-1}{j}$ for 0≤j<n yields the desired result.

Previous relation and Lemma 1 yields the sum of probability sums at the power of three:.

*Relation 7*
$$\sum_{|J|=j} P_J^3 = \sum_{l=1}^n p_l\left(\sum_{\substack{|J|=j \\ l \in J}} P_J^2\right) = \left(\sum_{l=1}^n p_l^3\right)\left(\binom{n-3}{j-1}-\binom{n-3}{j-2}\right)+3\left(\sum_{l=1}^n p_l^2\right)\binom{n-3}{j-2}+\binom{n-3}{j-3}$$

Applying this relation to uniform distributions produces the identity:

$$\binom{n}{j}\left(\frac{j}{n}\right)^3 = \frac{1}{n^2}\left(\binom{n-3}{j-1}-\binom{n-3}{j-2}\right)+\frac{3}{n}\binom{n-3}{j-2}+\binom{n-3}{j-3}.$$

*Relation 8 (Proof of Theorem 1)*

For all k, 0<k<n, there exists a set of coefficients $\alpha_{k,u}$, $1 \leq u \leq k$, such that $\sum_{|J|=j} P_J^k = \sum_{u=1}^{k} \binom{n-k}{j-u} \cdot \alpha_{k,u}$, where $\alpha_{k,u}$ are independent of j (i.e. depend only on n, k and the popularity) and $\alpha_{k,k} = 1$.

As shown previously, such a decomposition exists for k=1,2,3:

$$\sum_{|J|=j} P_J = \binom{n-1}{j-1} \; ; \; \sum_{|J|=j} P_J^2 = \binom{n-2}{j-2} + \binom{n-2}{j-1} \sum_{l=1}^{n} p_l^2 \; ; \; \sum_{|J|=j} P_J^3 = \binom{n-3}{j-3} + \binom{n-3}{j-2}\left(3\sum_{l=1}^{n} p_l^2 - \sum_{l=1}^{n} p_l^3\right) + \binom{n-3}{j-1}\sum_{l=1}^{n} p_l^3.$$

In the general case of exponent k and subset size j, we first prove the forthcoming relation 9, since from Relation 4 (ABS Lemma), for $1 \leq j \leq n$, $0 \leq k$:

$$\sum_{|J|=j} P_J^{k+1} = \sum_{l=1}^{n}\left(p_l \sum_{u=1}^{k+1} \binom{n-(k+1)}{j-u} \cdot \beta_{k+1,u}(p_l)\right) = \sum_{u=1}^{k+1}\left(\binom{n-(k+1)}{j-u} \cdot \sum_{l=1}^{n}(p_l \cdot \beta_{k+1,u}(p_l))\right)$$ which implies Relation 8, since obviously corresponding $\alpha_{k+1,u}$ are independent of j (i.e. depend only on n, k and the popularity distribution) and $\alpha_{k+1,k+1} = \sum_{l=1}^{n}(p_l \cdot \beta_{k+1,k+1}(p_l)) = \sum_{l=1}^{n} p_l = 1$.

Finally, let us note that, for k=2 and 3, it is true that: $\alpha_{k,k-1} = k - n + \sum_{l=1}^{n}(1-p_l)^k$, since

$$\sum_{l=1}^{n} p_l^2 = 2 - n + \sum_{l=1}^{n}(1-p_l)^2 \text{ and } 3\sum_{l=1}^{n} p_l^2 - \sum_{l=1}^{n} p_l^3 = 3 - n + \sum_{l=1}^{n}(1-p_l)^3.$$

*Relation 9*

For all k, 0<k<n-1, and any index l, $1 \leq l \leq n$, of the popularity distribution, there exists a set of coefficients $\beta_{k+1,u}(p_l)$, $1 \leq u \leq k+1$, such that $\sum_{\substack{|J|=j \\ l \in J}} P_J^k = \sum_{u=1}^{k+1} \binom{n-(k+1)}{j-u} \cdot \beta_{k+1,u}(p_l)$, where $\beta_{k+1,u}(p_l)$ is independent of j (i.e. depend on n, k and the popularity) and $\beta_{k+1,k+1}(p_l) = 1$.

As shown above, for k=1: $\sum_{\substack{|J|=j \\ l \in J}} P_J = p_l \binom{n-2}{j-1} + \binom{n-2}{j-2}$ and, k=2:

$$\sum_{\substack{|J|=j \\ l \in J}} P_J^2 = p_l^2 \binom{n-3}{j-1} + \left(2p_l - p_l^2 + \sum_{m=1}^{n} p_m^2\right)\binom{n-3}{j-2} + \binom{n-3}{j-3}.$$

In the general case of exponent k, we prove the dual relation:

$$\sum_{\substack{|J|=j \\ l \in J}} P_J{}^k + \sum_{\substack{|J|=j-1 \\ l \notin J}} P_J{}^k = \sum_{u=1}^{k+1} \binom{n-k}{j-u} \cdot \beta_{k+1,u}(p_l)$$ such that $\beta_{k+1,k+1}(p_l) = 1$. Note the only difference with

relation (9) is that the upper term of the binomial coefficient is n-k instead of n-(k+1).
Indeed, proving this relation is equivalent to proving relation (9) (under the same conditions).

The 'if' clause stems directly from identity $\binom{n-(k+1)}{j-u} + \binom{n-(k+1)}{j-1-u} = \binom{n-k}{j-u}$ and the 'only if' from identity $\sum_{i=1}^{j} \binom{n-k}{i-u}(-1)^i = (-1)^j \binom{n-k-1}{j-u}$, obtained by varying subset size i from j down to 1, multiplying by $(-1)^{i-j}$ and summing up the rows.

We proceed with a proof by strong induction on the exponent.

One has: $\sum_{\substack{|J|=j \\ l \in J}} P_J{}^k = \sum_{\substack{|J|=j-1 \\ l \notin J}} (P_J + p_l)^k = \sum_{\substack{|J|=j-1 \\ l \notin J}} \sum_{i=0}^{k} \binom{k}{i} p_l{}^{k-i} P_J{}^i = \sum_{i=0}^{k} \binom{k}{i} p_l{}^{k-i} \left( \sum_{|J|=j-1} P_J{}^i - \sum_{\substack{|J|=j-1 \\ l \in J}} P_J{}^i \right)$,

hence: $\sum_{\substack{|J|=j \\ l \in J}} P_J{}^k + \sum_{\substack{|J|=j-1 \\ l \notin J}} P_J{}^k = \sum_{i=0}^{k} \binom{k}{i} p_l{}^{k-i} \left( \sum_{|J|=j-1} P_J{}^i \right) - \sum_{i=0}^{k-1} \binom{k}{i} p_l{}^{k-i} \left( \sum_{\substack{|J|=j-1 \\ l \in J}} P_J{}^i \right)$.

Now, using Lemma 1, for 1≤j≤n, and i≥1: $\sum_{|J|=j-1} P_J{}^i = \sum_{m=1}^{n} \left( p_m \sum_{\substack{|J|=j-1 \\ m \in J}} P_J{}^{i-1} \right)$, which requires the

separate handling of the first value (i=0) of RHS first summation, one has:

$$\sum_{\substack{|J|=j \\ l \in J}} P_J{}^k + \sum_{\substack{|J|=j-1 \\ l \notin J}} P_J{}^k = p_l{}^k \binom{n}{j-1} + \sum_{i=1}^{k} \binom{k}{i} p_l{}^{k-i} \left( \sum_{m=1}^{n} \left( p_m \sum_{\substack{|J|=j-1 \\ m \in J}} P_J{}^{i-1} \right) \right) - \sum_{i=0}^{k-1} \binom{k}{i} p_l{}^{k-i} \left( \sum_{\substack{|J|=j-1 \\ l \in J}} P_J{}^i \right)$$

Changing the index of the first summation (decrementing)

$$\sum_{\substack{|J|=j \\ l \in J}} P_J{}^k + \sum_{\substack{|J|=j-1 \\ l \notin J}} P_J{}^k = p_l{}^k \binom{n}{j-1} + \sum_{i=0}^{k-1} \binom{k}{i+1} p_l{}^{k-i-1} \left( \sum_{m=1}^{n} \left( p_m \sum_{\substack{|J|=j-1 \\ m \in J}} P_J{}^i \right) \right) - \sum_{i=0}^{k-1} \binom{k}{i} p_l{}^{k-i} \left( \sum_{\substack{|J|=j-1 \\ l \in J}} P_J{}^i \right)$$

$$\sum_{\substack{|J|=j \\ l \in J}} P_J{}^k + \sum_{\substack{|J|=j-1 \\ l \notin J}} P_J{}^k = p_l{}^k \binom{n}{j-1} + \sum_{i=0}^{k-1} p_l{}^{k-i-1} \left( \binom{k}{i+1} \sum_{m=1}^{n} p_m \sum_{\substack{|J|=j-1 \\ m \in J}} P_J{}^i - \binom{k}{i} p_l \sum_{\substack{|J|=j-1 \\ l \in J}} P_J{}^i \right).$$

Although very cumbersome, this expression is interesting because exponents of $P_J$ in both summations of RHS now vary from 0 up to k-1, so we can assume induction hypothesis holds for any exponent i: 1≤i<k<n-1. Thus, there exists a set of coefficients $\beta_{i+1,u}(p_l)$, 1≤u≤i+1, such that

$$\sum_{\substack{|J|=j-1 \\ l \in J}} P_J{}^i = \sum_{u=1}^{i+1} \binom{n-(i+1)}{(j-1)-u} \cdot \beta_{i+1,u}(p_l),$$ where $\beta_{i+1,u}(p_l)$ are independent of j (i.e. depend only on n, i and popularity distribution) and $\beta_{i+1,i+1}(p_l) = 1$.

$$\sum_{\substack{|J|=j \\ l \in J}} P_J{}^k + \sum_{\substack{|J|=j-1 \\ l \in J}} P_J{}^k = p_l^k \binom{n}{j-1} + \sum_{i=0}^{k-1} p_l^{k-i-1}\left(\binom{k}{i+1}\sum_{m=1}^{n} p_m \sum_{u=1}^{i+1}\binom{n-(i+1)}{(j-1)-u}\cdot \beta_{i+1,u}(p_m) - \binom{k}{i}p_l \sum_{u=1}^{i+1}\binom{n-(i+1)}{(j-1)-u}\cdot \beta_{i+1,u}(p_l)\right).$$

or: $$\sum_{\substack{|J|=j \\ l \in J}} P_J{}^k + \sum_{\substack{|J|=j-1 \\ l \in J}} P_J{}^k = p_l^k \binom{n}{j-1} + \sum_{i=0}^{k-1}\sum_{u=1}^{i+1}\binom{n-(i+1)}{(j-1)-u}\left(\left(\binom{k}{i+1}\sum_{m=1}^{n} p_m \beta_{i+1,u}(p_m) - \binom{k}{i}p_l \beta_{i+1,u}(p_l)\right) p_l^{k-i-1}\right).$$

Let us denote $\gamma_{l,k}(i,u)$ the last expression in parenthesis (for given exponent k and element l, it depends only on indexes i and u): $$\sum_{\substack{|J|=j \\ l \in J}} P_J{}^k + \sum_{\substack{|J|=j-1 \\ l \in J}} P_J{}^k = p_l^k \binom{n}{j-1} + \sum_{i=0}^{k-1}\sum_{u=1}^{i+1}\binom{n-(i+1)}{(j-1)-u}\gamma_{l,k}(i,u).$$ The two RHS summations can be commuted: $$\sum_{\substack{|J|=j \\ l \in J}} P_J{}^k + \sum_{\substack{|J|=j-1 \\ l \in J}} P_J{}^k = p_l^k \binom{n}{j-1} + \sum_{u=1}^{k}\sum_{i=u-1}^{k-1}\binom{n-(i+1)}{(j-1)-u}\gamma_{l,k}(i,u)$$

We want to show that this expression can be written as a sum of binomial coefficients with weights independent of j: $$\sum_{\substack{|J|=j \\ l \in J}} P_J{}^k + \sum_{\substack{|J|=j-1 \\ l \in J}} P_J{}^k = \sum_{u=1}^{k+1}\binom{n-k}{j-u}\cdot \beta_{k+1,u}(p_l)$$ such that $\beta_{k+1,k+1}(p_l) = 1$.

We can transform the two binomial coefficients above by using a variant of Vandermonde identity which states that when $0 \leq j \leq n$ and $0 \leq k \leq n$: $\binom{n}{j} = \sum_{u=0}^{j}\binom{k}{u}\binom{n-k}{j-u}$. By noting that, if $j > k$, then for any u, $k < u \leq n$, the first binomial coefficient is null, and if $j < k$, then for any u, $j < u \leq k$, the second one is null, the upper bound of the summation index can be set to k: $\binom{n}{j} = \sum_{u=0}^{k}\binom{k}{u}\binom{n-k}{j-u}$.

Hence, previous expression can be written:
$$\sum_{\substack{|J|=j \\ l \in J}} P_J{}^k + \sum_{\substack{|J|=j-1 \\ l \in J}} P_J{}^k = p_l^k\left(\sum_{u=0}^{k}\binom{k}{u}\binom{n-k}{j-1-u}\right) + \sum_{u=1}^{k}\sum_{i=u-1}^{k-1}\binom{n-k}{v}\binom{k-1-i}{j-1-u-v}\gamma_{l,k}(i,u).$$

Obviously, the first part of the RHS has the desired form, the second one needs some restructuring which is obtained by commuting two summations:
$$\sum_{\substack{|J|=j \\ l \in J}} P_J{}^k + \sum_{\substack{|J|=j-1 \\ l \in J}} P_J{}^k = p_l^k\left(\sum_{u=1}^{k+1}\binom{k}{u-1}\binom{n-k}{j-u}\right) + \sum_{u=1}^{k}\sum_{v=0}^{n-k}\binom{n-k}{j-1-u-v}\left(\sum_{i=u-1}^{k-1}\binom{k-1-i}{v}\gamma_{l,k}(i,u)\right)$$

Then, it can be seen that when $u+v > k$, all the terms of the i-indexed summation are null (set $q = k-1-i$, all the binomial coefficients of the summation $\sum_{q=0}^{k-u}\binom{q}{v}\gamma_{l,k}(k-1-q,u)$ are null).

Hence for a given u ($1 \leq u \leq k$) v-indexed summation is non-null between 0 and k-u.

This leads to: $\sum_{\substack{|J|=j \\ l \in J}} P_J{}^k + \sum_{\substack{|J|=j-1 \\ l \in J}} P_J{}^k = \sum_{u=1}^{k+1}\binom{n-k}{j-u}\binom{k}{u-1}p_l^k + \sum_{u=1}^{k}\sum_{v=0}^{k-u}\binom{n-k}{j-1-u-v}\left(\sum_{i=u-1}^{k-1}\binom{k-1-i}{v}\gamma_{l,k}(i,u)\right).$

Changing variable v to w=v+u: $\sum_{u=1}^{k+1}\binom{n-k}{j-u}\binom{k}{u-1}p_l^k + \sum_{u=1}^{k}\sum_{w=u}^{k}\binom{n-k}{j-1-w}\left(\sum_{i=u-1}^{k-1}\binom{k-1-i}{w-u}\gamma_{l,k}(i,u)\right)$

Rearranging the double summation: $\sum_{u=1}^{k+1}\binom{n-k}{j-u}\binom{k}{u-1}p_l^k + \sum_{w=1}^{k}\binom{n-k}{j-1-w}\sum_{u=1}^{w}\left(\sum_{i=u-1}^{k-1}\binom{k-1-i}{w-u}\gamma_{l,k}(i,u)\right)$, which gives the desired decomposition. Note that obviously, weight $\beta_{1,k+1}(p_l)$ is equal to $p_l^k$.

We still need to show that the weight of $\binom{n-k}{j-(k+1)}$ in the expression, $\beta_{k+1,k+1}(p_l)$, is 1. It is equal to:

$p_l^k + \sum_{u=1}^{k}\left(\sum_{i=u-1}^{k-1}\binom{k-1-i}{k-u}\gamma_{l,k}(i,u)\right) = p_l^k + \sum_{u=1}^{k}\gamma_{l,k}(u-1,u)$, since binomial coefficient is non-null only for i=u-1. The second term is: $\sum_{u=1}^{k}\left(\binom{k}{u}\sum_{m=1}^{n}p_m\beta_{u,u}(p_m) - \binom{k}{u-1}p_l\beta_{u,u}(p_l)\right)p_l^{k-u}$ and since by induction hypothesis, $\beta_{u,u}=1$ for any u≤k, this is $\sum_{u=1}^{k}\left(\binom{k}{u} - \binom{k}{u-1}p_l\right)p_l^{k-u}$. It is then easy to show this expression reduces to $1 - p_l^k$ which, after adding $p_l^k$ from the first term, completes the proof.
QED.

## Appendix 3: Analytic Form of Decomposition for Uniform Distributions

For a uniform distribution, $\sum_{|J|=j} P_J^{\ k} = \binom{n}{j}\left(\dfrac{j}{n}\right)^k$, and k<n, an analytic form $\sum_{u=1}^{k}\binom{n-k}{j-u}\alpha_{k,u}$ can be obtained using Stirling numbers of the second order.

First, Chu-Vandermonde identity: $\sum_{j=0}^{k}\binom{m}{j}\binom{n-m}{k-j} = \binom{n}{k}$, which holds $\forall$ k, m integers $0 \le k \le n$, and $0 \le m \le n$, is used to generate a decomposition of $\binom{n-i}{j-i}$ in the form of a summation:

$\binom{n-i}{j-i} = \sum_{v=0}^{j-i}\binom{m}{v}\binom{n-i-m}{j-i-v}$ where $0 \le j-i \le n-i$, and $0 \le m \le n-i$. Then, introducing variable k=m+i

(hence $i \le k \le n$) gives: $\binom{n-i}{j-i} = \sum_{v=0}^{j-i}\binom{k-i}{v}\binom{n-k}{j-i-v}$, followed by variable u=v+i, which yields:

$\binom{n-i}{j-i} = \sum_{u=i}^{j}\binom{k-i}{u-i}\binom{n-k}{j-u}$. Note that summation index u can be extended to range [1..n], since additional terms are null (first binomial is null for u<i, and second for u>j).

Now, from the well-known relation of Stirling numbers, $j^k = \sum_{i=0}^{j} S(k,i)\binom{j}{i}i!$, one has:

$\binom{n}{j}j^k = \sum_{i=0}^{j} S(k,i)\binom{n}{j}\binom{j}{i}i! = \sum_{i=0}^{j} S(k,i)\binom{n}{i}\binom{n-i}{j-i}i!$. Note that summation index can be either j or k, since if j>k, S(k,i) is null for i>j and if j<k, the binomial coefficient is null for i>j.

With previous decomposition of the binomial coefficient, one has:

$\binom{n}{j}j^k = \sum_{i=0}^{j} S(k,i)\binom{n}{i}\left(\sum_{u=1}^{n}\binom{k-i}{u-i}\binom{n-k}{j-u}\right)i! = \sum_{u=1}^{n}\sum_{i=0}^{j} S(k,i)\binom{n}{i}\binom{k-i}{u-i}\binom{n-k}{j-u}i!$. Noting that, for $n \ge u > k$:

$\binom{k-i}{u-i} = 0$, and for u>j, $\binom{n-k}{j-u} = 0$ as well, sum can be replaced by: $\sum_{u=1}^{k}\binom{n-k}{j-u}\sum_{i=0}^{u} S(k,i)\binom{n}{i}i!\binom{k-i}{u-i}$

leading to (with index extended to n since u≤k<n): $\boxed{\alpha_{k,u} = \dfrac{1}{n^k}\sum_{i=1}^{n} S(k,i)i!\binom{k-i}{u-i}\binom{n}{i}}$.

It verifies for all n and k<n: $\alpha_{k,1}(n) = n^{1-k}$ and $\alpha_{k,k}(n) = \dfrac{1}{n^k}\sum_{i=0}^{n} S(k,i)i!\binom{n}{i} = 1$.

For k=3, one obtains: $\alpha_{3,2}(n) = \dfrac{3n-1}{n^2}$ which corresponds to an identity previously shown:

$$\binom{n}{j}\left(\dfrac{j}{n}\right)^3 = \dfrac{1}{n^2}\left(\binom{n-3}{j-1} - \binom{n-3}{j-2}\right) + \dfrac{3}{n}\binom{n-3}{j-2} + \binom{n-3}{j-3}.$$

For k=4, one obtains: $\alpha_{4,2}(n) = \dfrac{7n-4}{n^3}$ and $\alpha_{4,3}(n) = \dfrac{6n^2 - 4n + 1}{n^3}$.

Note that $\binom{n}{j} j^k = \sum_{u=1}^{k}\binom{n-k}{j-u}\sum_{i=1}^{u} S(k,i)\, i!\, \binom{k-i}{u-i}\binom{n}{i}$ could be readily extended to k=n. However this does not give a decomposition with the required condition because, in that case, weight is unique (u=j) and therefore not independent of j, thus leading to a tautology. Equally, for k>n, there is no such decomposition since all binomial coefficients $\binom{n-k}{j-u}$ are null.

**Appendix 4: Analytic Form of Decomposition for a Generalized Popularity**

Marco Riedel gave a decomposition using generating functions, under the condition $j \geq k$, with a $\binom{n-u}{j-u}$, $1 \leq u \leq k$, basis of binomial coefficients: $\sum_{|J|=j} P_J^{\ k} = \sum_{u=1}^{k} \binom{n-u}{j-u} \cdot \varepsilon_{k,u}$, where the weights are

$$\boxed{\varepsilon_{k,u} = \sum_{q=1}^{k} (-1)^{u-q} \binom{n-q}{u-q} \sum_{|J|=q} P_J^{\ k}}.$$

Note that when $j=k$, it leads to a tautology.

This decomposition can be readily ported to a coefficient basis $\binom{n-k}{j-u}$, $1 \leq u \leq k$, using Chu-Vandermonde identity: $\binom{n-u}{j-u} = \sum_{i=0}^{k-u} \binom{n-k}{j-u-i}\binom{k-u}{i}$, $1 \leq u \leq k$. So: $\sum_{|J|=j} P_J^{\ k} = \sum_{u=1}^{k} \sum_{i=0}^{k-u} \binom{n-k}{j-u-i}\binom{k-u}{i} \cdot \varepsilon_{k,u}$.

Noting $v=u+i$ such that $1 \leq v \leq k$, $\sum_{|J|=j} P_J^{\ k} = \sum_{u=1}^{k} \sum_{v=u}^{k} \binom{n-k}{j-v}\binom{k-u}{k-v} \cdot \varepsilon_{k,u}$. Lower bound of second sum index can be extended to $v=1$ since for, $v<u$, the second binomial coefficient is null:

$\sum_{|J|=j} P_J^{\ k} = \sum_{u=1}^{k} \sum_{v=1}^{k} \binom{n-k}{j-v}\binom{k-u}{k-v} \cdot \varepsilon_{k,u}$ giving : $\sum_{|J|=j} P_J^{\ k} = \sum_{v=1}^{k} \binom{n-k}{j-v} \sum_{u=1}^{k} \binom{k-u}{k-v} \cdot \varepsilon_{k,u}$. Hence:

$$\boxed{\alpha_{k,v} = \sum_{u=1}^{k} \binom{k-u}{k-v} \varepsilon_{k,u}}.$$

This allows for a direct derivation for $\alpha_{k,v}$, $1 \leq v \leq k$, coefficients for a general distribution from MR formula: $\alpha_{k,v} = \sum_{u=1}^{k} \binom{k-u}{k-v} \sum_{q=1}^{k} \binom{n-q}{u-q} (-1)^{u-q} \sum_{|J|=q} P_J^{\ k} = \sum_{q=1}^{k} \left( \sum_{u=1}^{k} \binom{n-q}{u-q}\binom{k-u}{v-u} (-1)^{u-q} \right) \sum_{|J|=q} P_J^{\ k}$.

One can notice that the inner sum is null for $u<q$ and also for $u>v$. Hence range of the sum is $u:[q,v]$, in other words, inner sum is null when $q>v$. Then, with index change $r=u-q$ and under the constraint $n-q>k-q \geq v-q$, the inner sum is: $\sum_{r=0}^{v-q} \binom{n-q}{r}\binom{k-q-r}{v-q-r} (-1)^r = (-1)^{v-q} \binom{n-k-1+v-q}{v-q}$.

Finally, since the inner sum is null for $q>v$, index upper bound can be replaced by $v$:

$$\boxed{\alpha_{k,v} = \sum_{q=1}^{v} (-1)^{v-q} \binom{n-k-1+v-q}{v-q} \sum_{|J|=q} P_J^{\ k}}$$

is the coefficient of the decomposition: $\sum_{|J|=j} P_J^{\ k} = \sum_{v=1}^{k} \binom{n-k}{j-v} \cdot \alpha_{k,v}$.

It can be checked that $\alpha_{k,1} = \sum_{|J|=1} P_J^{\ k} = \sum_{l=1}^{n} p_l^{\ k}$ and $\alpha_{k,k} = \sum_{q=1}^{k} (-1)^{k-q} \binom{n-1-q}{k-q} \sum_{|J|=q} P_J^{\ k}$ whose equality to 1 requires Corollary 2 to hold since it is a special case for $k=c-1$.

Also: $\sum_{|J|=n-1} P_J^k = \sum_{u=1}^{k} \binom{n-k}{n-1-u} \cdot \alpha_{k,u} = \alpha_{k,k-1} + (n-k)$, hence $\boxed{\alpha_{k,k-1} = k - n + \sum_{l=1}^{n}(1-p_l)^k}$.

It is not too difficult to show that for uniform distribution, Appendix 3 formula can be derived since: $\sum_{q=1}^{v}(-1)^{v-q}\binom{n-k-1+v-q}{v-q}\binom{n}{q}q^k$ is identical to $\sum_{i=1}^{n} S(k,i)i!\binom{k-i}{v-i}\binom{n}{i}$.

This can be checked by extending the Stirling number:

$$\sum_{i=1}^{n} S(k,i)i!\binom{k-i}{v-i}\binom{n}{i} = \sum_{i=1}^{n}\left(\sum_{q=0}^{i}(-1)^{i-q}\binom{i}{q}q^k\right)\binom{k-i}{v-i}\binom{n}{i} = \sum_{i=1}^{v}\sum_{q=1}^{i}(-1)^{i-q}q^k\binom{k-i}{v-i}\binom{n}{q}\binom{n-q}{i-q}$$

This expression is: $\sum_{q=1}^{v}\left(\sum_{i=q}^{v}(-1)^{i-q}\binom{k-i}{v-i}\binom{n-q}{i-q}\right)\binom{n}{q}q^k$, and, with q≤v≤k<n, since $\sum_{i=q}^{v}(-1)^{i-q}\binom{k-i}{v-i}\binom{n-q}{i-q} = (-1)^{v-q}\binom{n-k-1+v-q}{v-q}$, (relation obtained on previous page with r=i-q variable change), this lends the desired result.

**Appendix 5: Computation of Sums of Powers of Subset Probabilities for k<<j**

Remarkable is that, when the subset size j is large and the power exponent k is small (k<<j), this relation provides a faster computation of sums of powers of subset probabilities.
Sum of k-powers of large subsets can be obtained as a linear sum of k-powers of smaller subsets with size at most equal to k.

From $\sum_{|J|=j} P_J^k = \sum_{u=1}^{k} \binom{n-k}{j-u} \alpha_{k,u}$, $0 \le k < j \le n$, with $\alpha_{k,u} = \sum_{q=1}^{u} (-1)^{u-q} \binom{n-k-1+u-q}{u-q} \sum_{|J|=q} P_J^k$, an analytic expression of the j-size summation as a sum of q-size summations for q from 1 to k can be obtained: $\sum_{|J|=j} P_J^k = \sum_{q=1}^{k} \eta_{k,j}(q) \sum_{|J|=q} P_J^k$ with $\eta_{k,j}(q) = \sum_{u=q}^{k} \binom{n-k}{j-u}(-1)^{u-q} \binom{n-k-1+u-q}{u-q}$, by permutation of the summations.

Note that $\eta_{k,j}(j) = 1$, and $\eta_{k,j}(q) = 0$, for $j < q \le k$, and:
$$\eta_{k,j}(q) = \sum_{u=0}^{k-q} \binom{n-k}{j-q-u}(-1)^u \binom{n-k-1+u}{u} = (-1)^{k-q} \binom{n-k}{j-k}\binom{n-q}{n-k}\frac{j-k}{j-q}, \quad \text{for } 1 \le q < j.$$

It stands also that: $\eta_{k,k}(q) = 0$, for $q \ne k$.

In other words, when j>k, and particularly for large j and small k, we obtain a more efficient computation of the expression:
$$\boxed{\sum_{|J|=j} P_J^k = \binom{n-k}{j-k}(j-k)\sum_{q=1}^{k} \binom{n-q}{n-k}\frac{(-1)^{k-q}}{j-q} \sum_{|J|=q} P_J^k}.$$

For n=100, j=50, and k=5: the number of j-subsets to analyze compared to the number of 1..k-subsets is 1.27106×10^21 times larger !!!